\pdfoutput=1
\documentclass[nofootinbib,twocolumn, prd, preprintnumbers]{revtex4-1}
\usepackage{amsmath,amssymb,graphicx,bm,psfrag,color,slashed,subfigure,mhchem,array}

\usepackage[dvipsnames]{xcolor}

\usepackage{siunitx} 
\DeclareSIUnit{\year}{y}

\begin{document}

\preprint{IPPP/22/45, TTP22-043}
\vspace{0.5cm}
\title{\boldmath
The 
Axion-Higgs Portal}

\vspace{0.1cm}

\author{Martin Bauer$^a$}
\author{Guillaume Rostagni$^a$}
\author{Jonas Spinner$^{a,b}$}

\address{$^a$Institute for Particle Physics Phenomenology, Department of Physics\\
Durham University, Durham, DH1 3LE, United Kingdom}

\address{$^b$Institut f\"ur Theoretische Teilchenphysik, Universit\"at Karlsruhe, Karlsruhe Institute of Technology, D-76128 Karlsruhe}

\begin{abstract}
The phenomenology of axions and axion-like particles strongly depends on their couplings to Standard Model particles. The focus of this paper is the phenomenology of the unique dimension six operator respecting the shift symmetry: the axion-Higgs portal. We compare constraints from Higgs physics, flavor violating and radiative meson decays, bounds from atomic spectroscopy searching for fifth forces and astrophysical observables. In contrast to the QCD axion, axions interacting through the axion-Higgs portal are stable and can provide a dark matter candidate for any axion mass. We derive the parameter space for which freeze-out and freeze-in production as well as the misalignment mechanism can reproduce the observed relic abundance and compare the results with the phenomenological constraints. For comparison we also derive  Higgs, flavor and spectroscopy constraints and the parameter space for which the scalar Higgs portal without derivative interactions can explain dark matter.
\end{abstract}

\maketitle

\section{Introduction}
Axions and axion-like particles are interesting candidates for new physics, because they are a generic feature of extensions of the Standard Model (SM) in which an approximate global symmetry is spontaneously broken. For the purpose of this paper we colloquially use \emph{axion} to refer to this class of particles.
Interactions between SM particles and axions are described by an effective field theory with operators of mass dimension $\geq 5$ and suppressed by the scale $f$ at which the global symmetry is broken. For example the interaction of an axion $a$ to gluons is described by the operator
\begin{align}\label{eq:QCDaxion}
\mathcal{L}^5_a \supset \frac{a}{f}G^{\mu\nu}_a\tilde G_{\mu\nu}^a \,,
\end{align}
where $\tilde G_{\mu\nu}=\frac{1}{2}\epsilon_{\mu\nu\rho\sigma}G^{\rho\sigma}$ is the dual of the QCD field strength tensor. If such a coupling to gluons is present, the axion can explain why QCD seems to preserve CP~\cite{Peccei:1977hh,Peccei:1977ur,Weinberg:1977ma,Wilczek:1977pj}. The most general set of dimension 5 operators describing axion interactions with SM particles also includes couplings to electroweak gauge bosons and SM fermions~\cite{Georgi:1986df}. The mass term explicitly breaks the shift symmetry and for sufficiently small masses, axions are long-lived and can contribute to the observed relic abundance of dark matter in the universe~\cite{Preskill:1982cy}.  

An interaction between the Higgs boson, the $Z$ boson and an axion is only present at dimension 5 if it is induced by axion couplings to chiral fermions~\cite{Bauer:2017nlg,Bauer:2017ris}. In contrast, the axion-Higgs portal (or \emph{derivative} Higgs portal) is a dimension six operator~\cite{Weinberg:2013kea}
\begin{align}\label{eq:ALPHiggsportal}
   \mathcal{L}^6_a = \frac{c_{ah}}{f^2}\partial_\mu a \partial^\mu a\, \phi^\dagger \phi\,,
\end{align}
where $\phi$ denotes the Higgs doublet. Importantly, the operator \eqref{eq:ALPHiggsportal} is the leading operator in the $1/f$ expansion that is bilinear in the axion fields and invariant under the $Z_2$ transformation $a\to -a$. In this paper we consider this $Z_2$ symmetry as a consequence of the UV completion which remains unbroken at the level of the effective theory. Such an axion has several interesting features. It can be a dark matter candidate independent of its mass. Interactions with light SM particles are strongly suppressed by Higgs couplings as well as by the momentum suppression due to the two derivatives in \eqref{eq:ALPHiggsportal}. As a consequence it is particularly challenging to discover an axion that interacts with the SM through the operator \eqref{eq:ALPHiggsportal} and observables based on very precise measurements at low energies can be less sensitive compared to high-energy probes. Here we will discuss the phenomenology of an axion with a $Z_2$ symmetry and a softly broken shift-symmetry. We compare constraints from spectroscopy experiments, flavor violating and flavor conserving meson decays, invisible Higgs decays and astrophysics. We further calculate the parameter space for which such an axion can contribute to the observed relic abundance of dark matter in the Universe, comparing misalignment, freeze-in and freeze-out mechanisms. The results are compared with the renormalizable Higgs portal model in which a scalar field is invariant under a $Z_2$ transformation, but not under a shift symmetry. Throughout the paper we compare these results with the corresponding parameter space for a scalar Higgs portal 
\begin{align}
    \mathcal{L}\ni c_{sh} s^2 \phi^\dagger \phi,
    \label{eq:higgsPortal}
\end{align}
which is not protected by a shift symmetry, but invariant under a $Z_2$ transformation $s\to -s$. Therefore --in contrast to the axion-- the mass of the scalar $s$ is not protected against quadratically divergent radiative corrections.

\section{The axion-Higgs portal at different scales}\label{sec:aHportal}
Below the scale $\mu=f$ the theory of the axion-Higgs portal is defined by an extension of the SM with a real spin-0 field $a$ and an effective Lagrangian
\begin{align}\label{eq:EFTLag}
\mathcal{L}(\mu <f)=\frac{1}{2}\partial_\mu a\partial^\mu \!a -\frac{1}{2}m_a^2 a^2 +\frac{c_{ah}}{f^2}\partial_\mu a\partial^\mu \!a\phi^\dagger \phi +\mathcal{L}_\text{SM}\,,
\end{align}
where $\mathcal{L}_\text{SM}$ denotes the SM Lagrangian. For a vanishing axion mass $m_a$, this Lagrangian is invariant under a shift symmetry $a\to a+c$, where $c$ is a constant, and a $Z_2$ symmetry $a\to -a$.
The $Z_2$ symmetry implies that the effects of the axion do not introduce any additional parity violation, independent of it being a scalar or a pseudoscalar particle. The discussion in this paper does not depend on the specific UV completion of~\eqref{eq:EFTLag}, but we introduce a particularly simple UV completion of the SM with only one complex scalar field in Appendix~\ref{sec:UVcompletion}. 
Below the electroweak scale $\mu<v$ we integrate out the $W^\pm$ and $Z$ gauge bosons as well as the Higgs scalar and the top quark, so that we can write the effective Lagrangian as  
\begin{align}\label{eq:EFT}
    \mathcal{L}(\mu<v) = &
 \frac{c_{ah}c_\gamma}{f^2 m_h^2} (\partial_\mu a)^2 F_{\rho\sigma} F^{\rho\sigma} +\frac{c_{ah} c_G}{f^2 m_h^2}  (\partial_\mu a)^2 G_{\rho\sigma}^a G^{\rho\sigma}_a\notag\\[2pt]
 -\sum_{i,j} &\frac{c_{ah}c_{ij}}{f^2 m_h^2}  (\partial_\mu a)^2 \bar \psi_i \big(m_i P_L + m_j P_R \big) \psi_j\!+\!\text{h.c.}
\end{align}
The field strength tensors for photons and gluons are denoted by $F_{\rho\sigma}$ and $G_{\rho \sigma}^a$, respectively. Chiral projectors are defined as $P_{L/R}=1/2(1\mp \gamma_5)$, $m_i$ and $ m_j$ are the masses of the fermions $\psi_i$ and $\psi_j$ and the sum over $i,j=e,\mu,\tau, \nu_e, \nu_\mu, \nu_\tau, d,s,b,u,c$ extends over leptons and quarks apart from the top quark.  
The dimensionless Wilson coefficients are obtained to leading order in $m_h^2/(2m_t)^2$ and $m_h^2/(2m_W)^2$. 
In this limit the couplings between the axion and gauge bosons read~\cite{Kniehl:1995tn}
\begin{align}
 c_\gamma &=   -\frac{\alpha}{4\pi}\frac{47}{18}\,,\qquad 
  c_G =  \frac{\alpha_s}{4\pi}\frac{1}{3}\,,
\end{align}
and the couplings of the axion to SM fermions are given by
\begin{align}\label{eq:FDcouplings}
   c_{ii}=  \frac{1}{2}\,,
\end{align}
in the case of flavor diagonal couplings. Flavor-violating axion couplings are induced through the Higgs penguin~\cite{Grzadkowski:1983yp,Dedes:2003kp,Kachanovich:2020yhi} and are only relevant for external down-type quarks. For example the axion-coupling to $d$ and $s$ quarks is given by
\begin{align}\label{eq:cds}
c_{ds}= -\frac{3}{16\pi^2} \frac{m_t^2}{v^2} V_{td}^* V_{ts}\,.
\end{align}
The Wilson coefficients for the flavor changing transitions $b\to d$ and $b\to s$ can be obtained by replacing the CKM elements in \eqref{eq:cds}. Flavor changing axion couplings to up-type quarks are suppressed by $m_b^2/v^2$ at the amplitude level and charged lepton flavor-changing couplings are suppressed by neutrino masses. 

In contrast to the QCD axion or more generally axions that interact linearly with quarks or gluons~\cite{Bauer:2021wjo}, the axion-Higgs portal does not induce mixing between the neutral pion or other pseudoscalar mesons with the axion as long as the $Z_2$ symmetry remains unbroken. 
At energies below the QCD scale $\Lambda_\text{QCD}$, the relevant degrees of freedom are nucleons, leptons and photons. The effective Lagrangian for interactions induced by the axion-Higgs portal read then
\begin{align}
\mathcal{L}(\mu<\Lambda_\text{QCD}) =& \frac{c_{ah}c_N}{f^2m_h^2}m_N (\partial_\mu a)^2\bar N N +    \frac{c_{ah}m_\ell}{f^2m_h^2}(\partial_\mu a)^2\bar \ell \ell\notag\\
&+\frac{c_{ah}c_\gamma}{f^2 m_h^2} (\partial_\mu a)^2 F_{\rho\sigma} F^{\rho\sigma}\,,
\end{align}
where the nucleons are protons and neutrons $N=p,n$ with mass $m_N$ and the leptons can be electrons or muons $\ell=e,\mu$ with mass $m_\ell$. The coupling to nucleons can be written as
\begin{align}
c_N=\sum_{q=u,d,s} f_q^N + \frac{6}{27}f_{TG} \,,    
\end{align}
with the matrix elements defined by 
\begin{align}
 f_q^N\equiv \frac{\langle N | m_q \bar q q| N \rangle}{m_N}\,,\quad \frac{8\pi}{9\alpha_s}f_{TG}=-\frac{\langle N|G_{\rho\sigma}^a G^{\rho\sigma}_a |N\rangle}{m_N}\,, 
\end{align}
that can be determined from pion-nucleon scattering~\cite{Alarcon:2011zs,Crivellin:2013ipa, Hoferichter:2015dsa}. Using the results from \cite{Arcadi:2019lka}, we find the numerical values
\begin{align}
c_p\approx c_n \approx 0.30\,.   
\end{align}
We will neglect the mass difference between the proton and the neutron.

\section{Phenomenological Constraints on the axion-Higgs Portal\label{sec:higgsMeson}}

\begin{figure} 
    \centering
    \includegraphics[scale=.35]{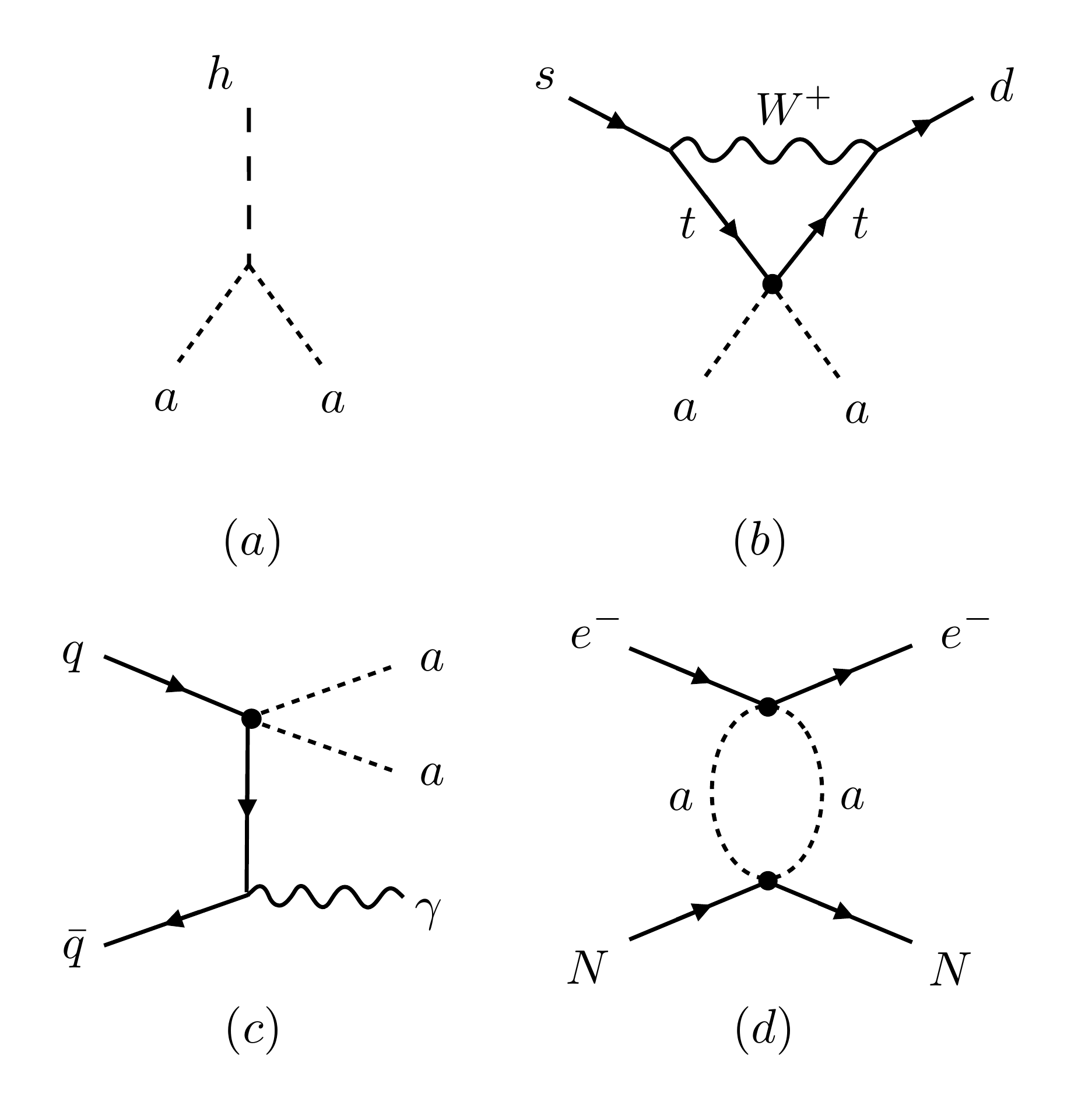}
    \caption{Diagrams for different processes induced by the axion-Higgs portal: (a) the Higgs decay into two axions, (b) a contribution to the flavor changing transition $s\to d a a$, (c) a contribution to the vector meson annihilation $V \to \gamma a a$, and (d) a contribution to the potential between electrons and nuclei generated by the exchange of axion pairs..}
    \label{fig:diagrams}
\end{figure}

After the preparations in the last section, we are now ready to calculate the predictions of the axion-Higgs portal and compare to experimental data. The phenomenology of the axion-Higgs portal is different from the QCD axion and other linearly coupled axion-like particles. Axions and ALPs are constrained by searches for the direct production at colliders~\cite{Bauer:2017ris,Jaeckel:2015jla,Alonso-Alvarez:2018irt} and indirect effects in lab based experiments such as light-shining through the wall experiments~\cite{Redondo:2010dp,OSQAR:2015qdv} or cavity resonance searches~\cite{Hagmann:1989hu} and astrophysical observables~\cite{Raffelt:2006cw}. Many of these experiments are only sensitive to axial interactions or axions decaying into SM final states, which are both absent in the case of the axion-Higgs portal. In the following we present existing bounds on the ratio $c_{ah}/f^2$ and discuss the best experimental strategy to discover a sterile axion.

\subsection{Higgs decays} 
The study of Higgs decays is the most direct way to probe the axion-Higgs portal.  The corresponding diagram is shown in Figure~\ref{fig:diagrams} (a) and the decay rate reads 
\begin{align}\label{eq:higgsDecay}
    \Gamma(h\to aa) = \frac{v^2 m_h^3}{32\pi} \frac{c_{ah}^2}{f^4} \bigg( 1-\frac{2m_a^2}{m_h^2}\bigg)^2 \sqrt{1-\frac{4m_a^2}{m_h^2}}.
\end{align}
In absence of linear interactions the axion is stable, leading to invisible Higgs decays.
Bounds on the branching ratio are set by searches for invisible decays of Higgs bosons produced in vector-boson fusion $\mathcal{B}(h\to \text{inv}) \le 0.145$ at ATLAS~\cite{ATLAS:2022yvh} and $\mathcal{B}(h\to \text{inv}) \le 0.18 $  at CMS~\cite{CMS:2022qva} at 95\% CL. Global fits result in slightly stronger bounds of $\mathcal{B}(h\to \text{inv}) \le 0.13$ at ATLAS~\cite{ATLAS:2022vkf} and $\mathcal{B}(h\to \text{inv}) \le 0.16 $  at CMS ~\cite{ CMS:2022dwd}. The reach of the high luminosity LHC and potential future colliders is given in Table~\ref{tab:HtoInv_projections}. 
\begin{table}
    \centering
    \begin{tabular}{p{2.1cm} p{2.0cm} p{2.1cm} c}
    \toprule 
    experiment & $\mathcal{B}(h\to \text{inv})
    $ & $c_{ah}/f^2 [\si{GeV^{-2}}]$ & ref \\
    \colrule
    LHC (today) & $1.45\times 10^{-1}$ & $7.1\times 10^{-7}$ 
    & \cite{ATLAS:2022yvh} \\
    HL-LHC & $2\times 10^{-2}$  & $2.6\times 10^{-7}$
    & \cite{Cepeda:2019klc} \\
    ILC \SI{250}{GeV} & $4.4\times 10^{-3}$ & $1.2\times 10^{-7}$
    & \cite{Baer:2013cma} \\
    FCC-hh & $2.5\times 10^{-4}$ & $3\times 10^{-8}$ 
    & \cite{FCC:2018vvp} \\
    \botrule
    \end{tabular}
    \caption{Current limits and projections for experimental bounds on the branching ratio of Higgs bosons to invisible final states.}
    \label{tab:HtoInv_projections}
\end{table}

\begin{table}
    \centering
    \begin{tabular}{   p{3cm} p{2.cm} p{2cm} c}
    \toprule 
    Decay width &upper bound& $c_{ah}/f^2 [\si{GeV^{-2}}]$ & ref \\
    \colrule
    $\mathcal{B}(K^+\to \pi^+ +\text{inv})$ & $4.8\times 10^{-11}$ &11.5& \cite{NA62:2020pwi, NA62:2021zjw}\\
    $\mathcal{B}(B^+\to K^+ +\text{inv})$  & $1.6\times 10^{-5}$ &$6.0\times 10^{-2}$& \cite{BaBar:2013npw} \\
    $\mathcal{B}(B^+\to \pi^+ +\text{inv})$ & $1.4\times 10^{-5}$& $2.3\times 10^{-1}$& \cite{Belle:2017oht} \\
    $\mathcal{B} (B^0 \to \text{inv})$ & $2.4\times 10^{-5}$ &$2.0\times 10^{-1}$& \cite{BaBar:2012yut} \\
    $\mathcal{R}(\Upsilon\to \gamma +\text{inv})$ & $3.5\times 10^{-6}$&$1.4\times10^{-1}$ & \cite{Belle:2018pzt} \\
    \botrule
    \end{tabular}
    \caption{Experimental bounds on meson decays into final states with invisible particles. There are dedicated searches for $K^+\to \pi^++\text{inv}$ and $\Upsilon (1S)\to \gamma aa$ which yield $m_a$-dependent upper bounds, we list their bounds for small $m_a$ in this table.}
    \label{tab:Mesondecays}
\end{table}

\begin{figure*}
    \centering
       \includegraphics[scale=.58]{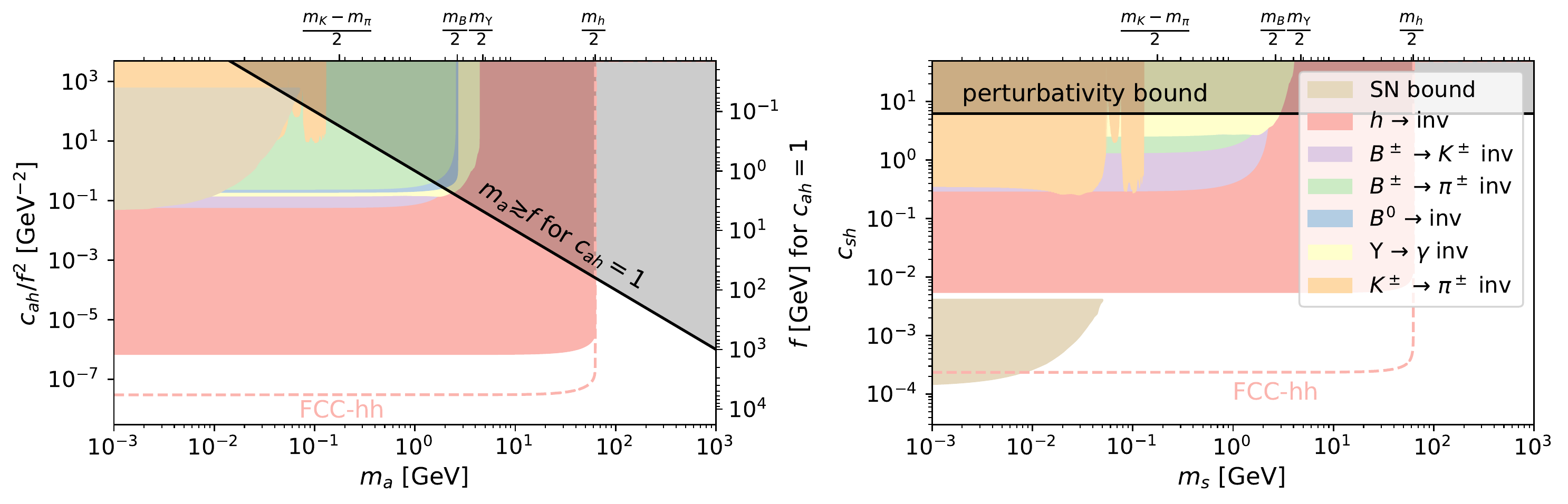}
    \caption{Constraints and projections from Higgs and flavor-violating meson decays and bounds from supernova energy loss for the axion-Higgs portal (left) and the Higgs portal (right). The color coding is indicated in the right panel. For the parameter space above the black solid line in the left panel the approximate shift symmetry is not a good assumption anymore, whereas the region above the black line in the right panel violates perturbativity. Note that these bounds are also relevant for lower values of $m_a$ for the case of the axion-Higgs portal, whereas cosmological constraints become stronger for the Higgs portal, see also Figure~\ref{fig:DMplot}. The supernova bound is taken from \cite{Bauer:2020nld}.}
    \label{fig:excPlot}
\end{figure*}

\subsection{Flavor-violating meson decays\label{sec:3bodyMeson}}

Some of the most sensitive probes of axions or ALPs in the mass range $m_a = 1-100$ MeV are meson decays like $K^+\to \pi^+ a$~\cite{Georgi:1986df, Bauer:2021mvw, Goudzovski:2022vbt}. These decays are forbidden for the axion-Higgs portal because they violate the $Z_2$ symmetry. Instead, axions interacting through the axion-Higgs portal are pair-produced in meson decays such as $K^+\to \pi^+ a a$ induced by diagrams like the one shown in Figure~\ref{fig:diagrams} (b). The 3-body phase space leads to a strong suppression of the decay rate 
\begin{align}
\Gamma(K^+\to \pi^+ a a) = \frac{m_{K^+}^9}{3\cdot 2^{13}\pi^3}\frac{c_{ah}^2}{f^4}\frac{c_{ds}^2}{m_h^4} F \bigg(\frac{m_a^2}{m_{K^+}^2},\frac{m_{\pi^+}^2}{m_{K^+}^2}\bigg),
\end{align}
where $c_{ds}$ is given in \eqref{eq:cds} and the function
\begin{align}
\begin{split}
    F (a,b) = 24(1-b)^2\int_{4a}^{(1-\sqrt{b})^2} \!dx\, (x-2a)^2 \\
    \times \sqrt{x-4a}\sqrt{\Big( \frac{1-b-x}{2\sqrt{x}}\Big)^2-b },
\end{split}
\end{align}
includes the phase space factor.\footnote{We find $F(0, m_\pi^2/m_K^2) = 0.172$, $F(0, m_\pi^2/m_B^2)=0.957$, $F (0, m_K^2/m_B^2)=0.722.$} Similar expressions hold for 3-body decays of $B$ mesons. Experimental bounds are given in Table~\ref{tab:Mesondecays}.

In addition to 3-body decays, the axion-Higgs portal predicts flavor-violating decays of neutral mesons to invisible final states. Similar to invisible Higgs decays it is experimentally very challenging to constrain invisible meson decays unless the meson recoils against SM particles. In $B$ factories the invisible decay of $B^0$ mesons can be observed through $e^+ e^- \to \Upsilon \to B^0\bar B^0$ with a subsequent invisible $B^0$ decay by tagging the second $B^0$ meson~\cite{BaBar:2012yut}. The decay rate reads
\begin{align}
\begin{split}
\Gamma (B^0\to aa) = &\frac{m_{B^0}^7}{128\pi} \frac{c_{ah}^2}{f^4} \frac{c_{bd}^2 f_{B^0}^2}{m_h^4} \bigg( 1-\frac{2m_a^2}{m_{B^0}^2}\bigg)^2\sqrt{1-\frac{4m_a^2}{m_{B^0}^2}},
\end{split}
\end{align}
where $f_{B^0}=\SI{190.5}{MeV}$~\cite{TUMQCD:2018fsq} is the neutral $B$ meson decay constant.

\subsection{Radiative Vector meson decays\label{sec:4VMeson}}

The axion-Higgs portal mediates the flavor-conserving vector meson decays $V \to \gamma a a$ via diagrams like the one shown in Figure \ref{fig:diagrams} (c). These decays are proportional to the flavor diagonal couplings \eqref{eq:FDcouplings}, which can be many orders of magnitude larger compared to the flavor violating couplings \eqref{eq:cds}. In line with the Wilczek equation~\cite{Wilczek:1977zn}, we use the ratio of decay widths 
\begin{align}
\mathcal{R}(V\to\gamma a a)&\equiv
    \frac{\Gamma(V\to\gamma a a)}{\Gamma(V \to e^+e^-)}\notag\\
    =\frac{1 }{3\cdot 2^{10}\pi^3 \alpha}&\frac{c_{ah}^2}{f^4}   \frac{m_V^8}{m_h^4} F\left(\frac{m_a^2}{M_V^2},0 \right) 
 \bigg[1-\frac{4\alpha_s}{3\pi}a_H(z)\bigg],
\end{align}
not including contributions suppressed by $\mathcal{O}(c_\gamma)$. We find the analytic result 
\begin{align}
\begin{split}
     F(x, 0) = &\, \sqrt{1 - 4x} \left( 1 - 10x + 42x^2 + 12x^3 \right) \\
    &- 24x^3(2-x) \log \frac{1+\sqrt{1-4x}}{1-\sqrt{1-4x}} \,,   
\end{split}
\end{align}
and use the NLO corrections $a_H(z)$ where $z=1-4m_a^2/m_V^2$ as given in~\cite{Nason:1986tr} such that $a_H(1)\approx 10$ for $m_a=0$ and $a_H(z)\propto z^{-1/2}$ in the limit $m_a\to m_V/2$.
The strongest constraints are currently set by BESIII~\cite{Liu:2022sdz} for $J/\psi$ decays and Belle~\cite{Belle:2018pzt} for $\Upsilon(1S)$ decays. The decay $V \to aa $ is forbidden, because two identical particles cannot be in an anti-symmetric spin 1 state. 

\begin{figure*}
        \centering
        \includegraphics[scale=.58]{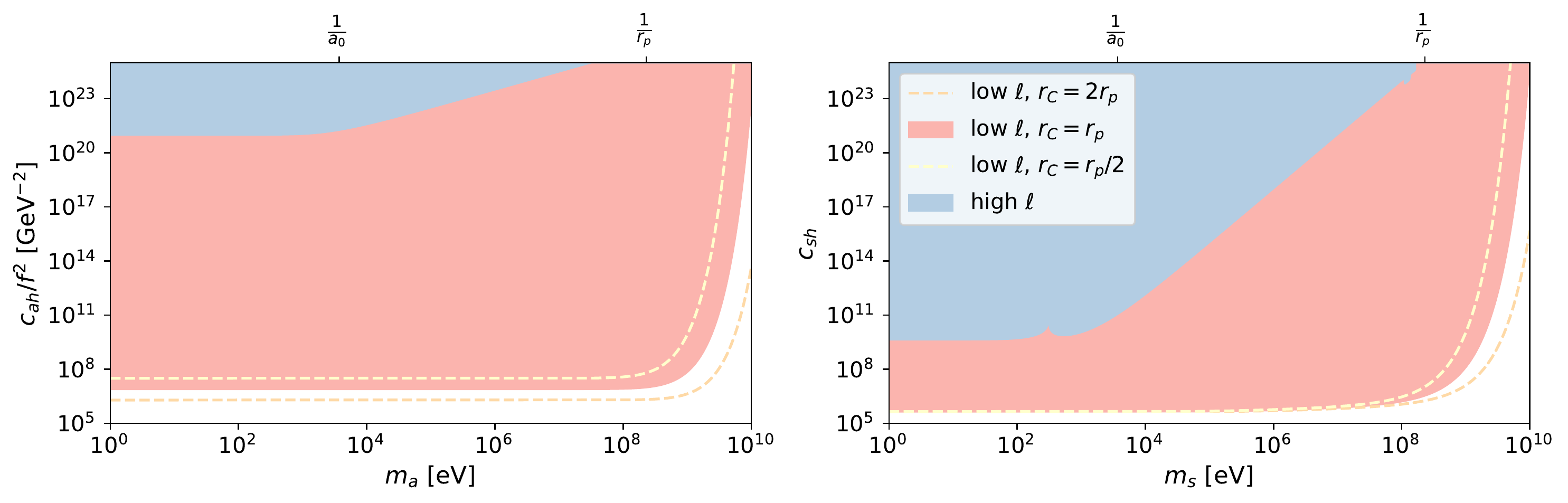}
    \caption{Bounds on the axion coupling and scale obtained from spectroscopic data for the axion-Higgs (left) and scalar Higgs (right) portals. The red bound is obtained from low-$\ell$ hydrogen states with a cutoff $r_C = r_p$; the dashed lines show the dependence of this bound on the chosen cutoff. In blue are the cutoff-independent bounds obtained from $\ell = 3$ hydrogen f-states.}
    \label{fig:ASplot}
\end{figure*}

\subsection{Constraints from atomic spectroscopy \label{sec:spectr}}

The exchange of pairs of axions as shown in Figure~\ref{fig:diagrams}~(d) induces a fifth force. This interaction through the axion-Higgs portal is strongly suppressed by two effective vertices proportional to  the inverse Higgs mass squared as well as to the small Higgs Yukawa couplings to stable SM particles.

The energy level shift in atoms due to this new interaction can be obtained by calculating the expectation value of the corresponding potential with respect to an electron in a given state
\begin{equation} \label{eq:energy-shift}
    \Delta E_{n\ell} = \langle \psi_{n\ell} | V(\mathbf r) | \psi_{n\ell} \rangle \equiv
    \int \mathrm d^3 \mathbf r\ |\psi_{n\ell}(\mathbf r)|^2 V(\mathbf r)
\end{equation}
where $\psi_{n\ell}(\mathbf r)$ is the wavefunction for the $(n\ell )$ state. For a radially symmetric potential, only the radial component of the wavefunction will enter the calculation and therefore the energy shifts for this potential will not depend on $m$. The potential can be obtained by taking the discontinuities in the scattering amplitude in the non-relativistic limit and perform the Fourier transform as a complex integration~\cite{Fichet:2017bng,Brax:2017xho}, 
\begin{align}\label{eq:Potaxion}
    V(r) = -  \frac{c_{ah}^2}{f^4}& \frac{c_p m_p m_e}{8\pi^3 m_h^4} \bigg[ \left( \frac{ m_a^5}{r^2} + \frac{15 m_a^3}{r^4} \right) K_1(2m_ar) \notag\\
   & + \left( \frac{6 m_a^4}{r^3} + \frac{30 m_a^2}{r^5} \right) K_2(2m_ar) \bigg]\,,    \end{align}
 where $K_\nu(2mr)$ are modified Bessel functions of the second kind. 
At short ranges (or low axion masses) they scale like a power law, meaning the potential is dominated by an $r^{-7}$ term, whereas they decay exponentially in the long range limit or large axion mass limit. This is a crucial difference between the Higgs portal and the exchange of two axion with linear interactions to SM fermions. In this case there are additional diagrams beyond Figure~\ref{fig:diagrams} (d) with internal fermion lines and the potential scales as $r^{-5}$ at short distances~\cite{Ferrer:1998ue}. 

Uncertainties can be minimised by taking the ratio of two different hydrogen transition, and currently the strongest constraint reads \cite{Endo:2012hp}, 
\begin{equation} \label{eq:states-ratio}
    \left. \frac{E_{2s_{1/2}} - E_{8d_{5/2}}}{E_{1s_{1/2}} - E_{3s_{1/2}}} \right|_\text{exp - SM} < (-0.5 \pm 3.1) \times 10^{-12},
\end{equation}
where the experimental result and theory predictions are taken from~\cite{Tiesinga:2021myr} and ~\cite{Yerokhin:2018gna}, respectively. The energy shift induced by the axion in hydrogen atoms is divergent for s-, p- and d-states due to the wavefunction decaying too slowly to compensate the $r^{-7}$ behavior of the potential at short ranges. We choose to regulate this divergence by introducing a cutoff scale $r_C$. 
Using a reference cutoff scale to be the proton radius $r_p = 0.84$ fm~\cite{Tiesinga:2021myr}, \eqref{eq:states-ratio} translates into a bound on the axion coupling at $1\sigma$ 
\begin{equation} \label{eq:low-l-bound}
    \frac{c_{ah}}{f^2} < 5 \times 10^{6} \left( \frac{r_{C}}{\text{fm}} \right)^2 \ \text{GeV}^{-2}\,,
\end{equation}
for $m_a \ll 1/r_p \approx 0.2$ GeV with a sharp drop in the constraint at higher masses. 
Note that this result strongly depends on the cutoff scale chosen and is therefore only to be taken as an estimate.

States with a higher-$\ell$ quantum number do not suffer from this divergence and we also consider the 4f, 5f, and 6f states of hydrogen, comparing measurements found in~\cite{NIST-ASdatabase} to SM values obtained using~\cite{Yerokhin:2018gna,NIST-Elevels,Drake:1990zz} to obtain the bound shown in blue in Figure~\ref{fig:ASplot}. For the 6f state of hydrogen no measurement exists and we assume an agreement between experiment and the SM similar to that obtained for 4f and 5f states measurements for the estimate shown in Figure~\ref{fig:ASplot}.

Molecular spectroscopy can yield stronger constraints at short distances in particular for systems in which an electron is replaced by a muon whose wavefunction has a larger overlap with the nucleus. It can also offer measurements precise enough to derive bounds on the model. Various systems are considered in~\cite{Brax:2017xho,Banks:2020gpu} with the strongest bound resulting from the binding energy of the $(\nu=1, J=0)$ state of the muonic molecular deuterium ion $dd\mu^+$ giving
\begin{equation}
    \frac{c_{ah}}{f^2} < 4.4 \times 10^{8} \left( \frac{r_{C}}{2.1\text{fm}} \right)^2 \text{GeV}^{-2}\,,
\end{equation}
for a cutoff set at the deuterium radius $r_d = 2.1$ fm~\cite{Tiesinga:2021myr} and assuming $c_d \approx c_p$ for the deuterium-Higgs coupling. This result is however also strongly dependent on the cutoff scale chosen as the full integral is divergent for this system~\cite{Brax:2017xho}. 
We contrast these constraints with the corresponding results for the Higgs portal scalar, the interaction term with scalar couplings gives rise to a different form for the potential
\begin{equation}\label{eq:Spot}
    V(r) = - \frac{1}{8\pi^3} c_{sh} c_p m_p m_e \frac{m_s}{r^2} K_1(2m_sr)\,.
\end{equation}
This potential scales as $r^{-3}$ in the short range limit yielding an approximately logarithmic dependence on the cutoff scale $r_C$. The results are shown in Figure~\ref{fig:ASplot}. The main difference between the two models is that for interactions induced by the axion-Higgs portal the bounds obtained from high-$\ell$ transitions are considerably weaker compared to transitions at low $\ell$, whereas the difference is not as extreme in the case of the scalar Higgs portal. This is due to the different scaling of the potentials \eqref{eq:Potaxion} and \eqref{eq:Spot} with $r$.
We also include the strongest molecular spectroscopy bounds from~\cite{Brax:2017xho,Banks:2020gpu} from the antiprotonic helium molecular ion $\bar p$He$^+$ 
\begin{equation}
    c_{sh} < 8.4 \times 10^4\,,
\end{equation}
for $m_s < 10^4$ eV, and from the $dd\mu^+$ ion with
\begin{equation}
    c_{sh} < 2.2 \times 10^5\,,
\end{equation}
for $m_s < 10^5$ eV. Note that the bounds on $c_{ah}/f^2$ in Figure~\ref{fig:ASplot} are so weak that for some values of $m_a$ contributions from the shift-symmetry breaking operator $m_a^2/f^2 a^2 \phi^\dagger \phi$ would generate the potential~\eqref{eq:Spot} for the axion-Higgs portal model with $c_{sh}=m_a^2/f^2$. In this case stronger constraints on $f$ can be extracted from the right panel of Figure~\ref{fig:ASplot}. 

\begin{figure*}
    \centering
    \includegraphics[scale=.58]{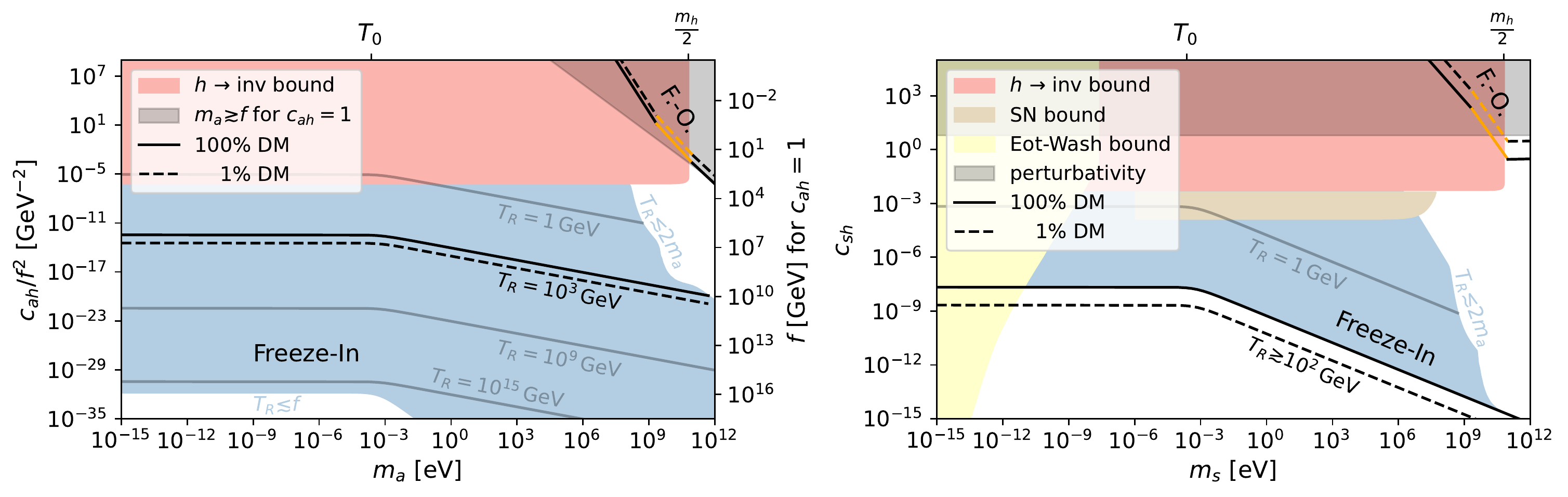}
    \caption{On the left, we show the required couplings for the axion-Higgs portal to explain the observed value for the energy density of dark matter $\Omega h^2 = 0.12~$\cite{Planck:2018vyg} for various mechanisms. On the right, we show the corresponding situation for the Higgs portal, the supernova and Eot-Wash bounds are taken from \cite{Bauer:2020nld}. The orange lines for freeze-out production are an estimate for a smooth intersection between the regimes that are dominated by $b\bar b\leftrightarrow aa$ and $\gamma\gamma\leftrightarrow aa$. For the case of freeze-in production, we show several lines to visualize the dependence on the cutoff temperature $T_R$. The blue regions visualize where freeze-in production is possible.}
    \label{fig:DMplot}
\end{figure*}

\subsection{Discussion}

The different constraints discussed in this section are shown in the $m_a - c_{ah}/f^2$ plane in the left panel of Figure~\ref{fig:excPlot}.
Given the hierarchy in precision between the constraints,
one would expect that low-energy experiments give stronger bounds for sufficiently small axion masses. However, due to the powerful double suppression by the axion derivative couplings and the factor $m_h^{-2}$ in the effective couplings, invisible Higgs boson decays result in the  strongest bound $c_{ah}/f^2\lesssim 10^{-6} \si{GeV^2}$ for the axion mass range where the decay is allowed. This bound is considerably stronger than bounds from searches for pair-produced axions in flavor violating or flavor conserving meson decays as well as the bound from supernova energy loss discussed in~\cite{Bauer:2020nld}. The constraints from atomic spectroscopy are so weak that they do not show up in Figure~\ref{fig:excPlot}. A dedicated plot is shown in Figure~\ref{fig:ASplot}. Note that for some of the values for $c_{ah}/f^2$ in Figure~\ref{fig:excPlot} and Figure~\ref{fig:ASplot} the bounds shown are only qualitative because the effective theory approach is not justified there. The parameter space shown in gray in the left panel of Figure~\ref{fig:excPlot} is excluded because the axion mass is larger than the decay constant and the assumption of an approximate shift symmetry is not justified. Axions with masses $m_a \geq m_h/2$ can only be produced in off-shell, invisible decays of the Higgs boson, which provide a significantly weaker bound~\cite{Ruhdorfer:2019utl,Argyropoulos:2021sav}. 
For comparison we show the constraints on a scalar Higgs portal with a stable scalar in the right panel of Figure~\ref{fig:excPlot} and Figure~\ref{fig:ASplot}, respectively. We derive the corresponding decay widths in Appendix~\ref{app:Higgsportal}. Here the strongest constraint for masses $m_s\gtrsim 50$ MeV is set by the bound on invisible Higgs decays, but for smaller scalar masses the constraint from supernova energy loss is stronger. Bounds from flavor constraints are stronger relative to the constraint from invisible Higgs decays if compared with the case of axion portal, because the scalar Higgs portal is renormalizable. These results are in stark contrast to axions with an approximate shift symmetry and linear couplings to Standard Model particles, for which flavor constraints are substantially stronger than the constraint from invisible Higgs decays~\cite{Bauer:2021mvw}.

\section{Dark matter through the axion-Higgs Portal}\label{sec:darkmatter}

 Linearly coupled axions such as the QCD axion can always decay into SM particles and need to be sufficiently long-lived to contribute to the observed dark matter abundance. In contrast, axions interacting with the SM through the axion-Higgs portal are stable independent of the axion mass and therefore provide a natural dark matter candidate.  Depending on the axion-Higgs portal coupling strength and the axion mass, the axions role in cosmology can be described by different simplifying assumptions, and the relic abundance of dark matter axions can be calculated using either freeze-out, freeze-in or vacuum misalignment mechanisms.

\subsection{Freeze-out production}

In the freeze-out scenario, axions are in equilibrium with the SM thermal bath until the equilibrium is lost due to the expansion of the universe.
The dominant interaction for freeze-out production of dark matter through the axion-Higgs portal depends sensitively on the axion mass. It will turn out that freeze-out production is only possible for couplings so large that they violate the consistency requirement $m_a \ll f$. We therefore only give a qualitative estimate, neglecting effects from the Higgs boson pole and particle thresholds. 

If freeze-out happens between big bang nucleosynthesis and the QCD phase transition $\SI{1}{MeV}\lesssim T\lesssim \SI{100}{MeV}$, the dominant process is photon-photon annihilation  $\gamma\gamma \leftrightarrow aa$ and the relic abundance is given by
\begin{align}
\begin{split}
    \frac{\Omega h^2}{0.12} &= 30 \frac{sh^2}{0.12\rho_c} \frac{f^4 m_h^4}{c_{ah}^2 c_\gamma^2 m_a^6 m_\text{Pl}} \\
    &\approx  3.10 \Big( \frac{m_a}{\SI{100}{MeV}}\Big)^{-6} \Big( \frac{c_{ah}/f^2}{10^6\si{GeV^{-2}}}\Big)^{-2}\,,
\end{split}
\end{align}
where $s$ is the entropy density today and $\rho_c$ is the critical energy density of the universe, $m_\text{Pl}$ is the Planck scale and $h$ is the reduced Hubble constant. The corresponding range of axion masses is $\SI{17}{MeV}\lesssim m_a \lesssim \SI{2.2}{GeV}$ and freeze-out happens at $x_\text{FO} = m_a/T_\text{FO} \approx 1.1\log (m_a / \SI{100}{MeV}) + 18.7$.
If freeze-out happens at temperatures above the bottom quark threshold $T \gtrsim m_b$, the dominant process is $b\bar b\leftrightarrow aa$ annihilation and we obtain for the correct relic abundance in this parameter space 
\begin{align}
\begin{split}
    \frac{\Omega h^2}{0.12} &= 6\sqrt{10} \frac{sh^2}{0.12\rho_c} \frac{f^4 m_h^4}{c_{ah}^2 m_b^2 m_a^4 m_\text{Pl}}\\
    &\approx 0.28 \Big( \frac{m_a}{10^2\si{GeV}}\Big)^{-4} \Big( \frac{c_{ah}/f^2}{10^{-4}\si{GeV^{-2}}}\Big)^{-2}.
\end{split}
\end{align}
The corresponding axion mass range is $\SI{95}{GeV}\lesssim m_a \lesssim \SI{2.2}{TeV}$ and freeze-out happens at $x_\text{FO} = m_a / T_\text{FO} \approx 1.1\log (m_a/\SI{100}{GeV}) + 23.7$. For intermediate axion masses, the dominant interaction for freeze-out production changes at the different fermion mass thresholds.

\subsection{Freeze-in production}
Freeze-in production occurs if the interaction strength between axions and SM particles is so small that equilibrium is never reached~\cite{Hall:2009bx}. Instead, axions are produced in the decays and scattering processes of SM particles until the SM particles go out of equilibrium.
The relic abundance from freeze-in production does in general depend on a cutoff temperature $T_R$, where the standard prescription of cosmology becomes invalid. This cutoff temperature is often interpreted as the reheating temperature, but we consider it the scale where our effective model becomes invalid. The earliest probe of cosmology comes from big bang nucleosynthesis, translating into a strong lower bound $T_R\gtrsim \SI{10}{MeV}$~\cite{Domcke:2015iaa}. A cutoff temperature around the electroweak scale would strongly constrain standard cosmology, but can not be ruled out from the current perspective. Regarding high values of $T_R$, the only strong upper bound in our model is $T_R \lesssim f$, because the prescription of the axion-Higgs portal as an effective field theory breaks down at that scale. Requiring that the axion-Higgs portal produces the observed amount of dark matter, this translates into $T_R \lesssim 8\times 10^{15}\,\si{GeV}$ for $m_a=0$ and even higher values for non-vanishing axion masses. For large axion masses there is another strong bound $T_R < 2 m_a$ from the requirement that thermal production of axions is kinematically allowed. Finally, the dependence on the cutoff temperature disappears in the limit of large $T_R$ and renormalizable couplings and is therefore unphysical. It turns out that this dependence is crucial for the phenomenology of the axion-Higgs portal. 

\begin{figure*}
    \centering
    \includegraphics[scale=.58]{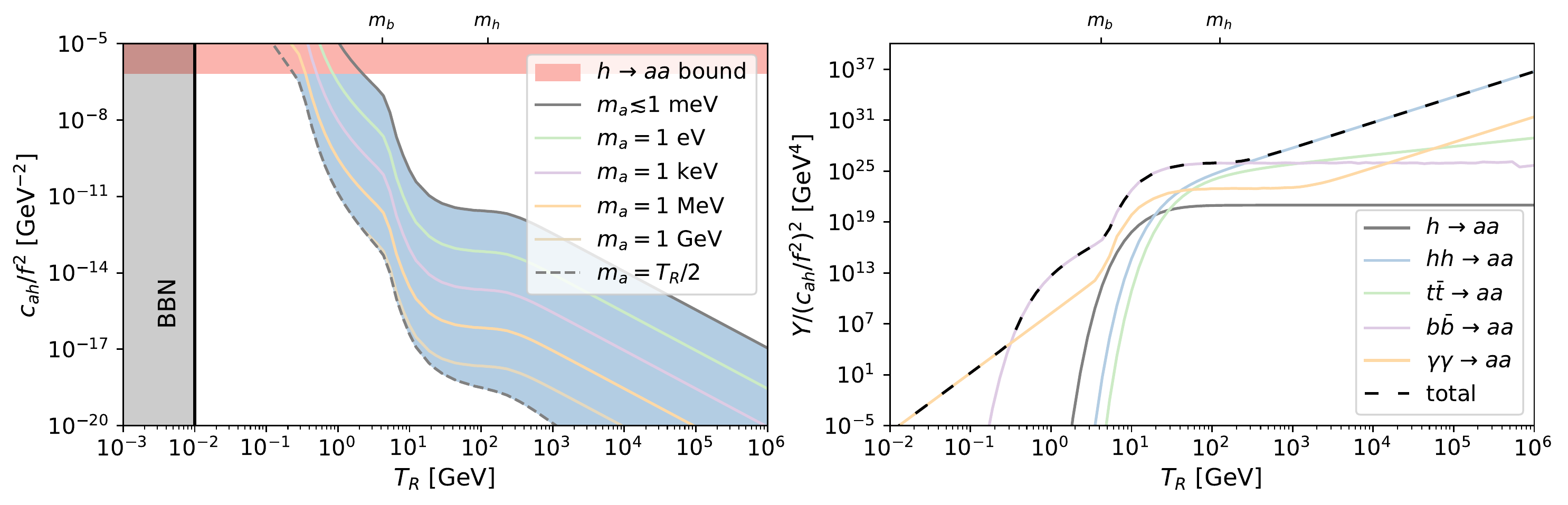}
    \caption{We visualize the effects of the dependence of freeze-in production on the cutoff temperature $T_R$. The left panel is complementary to the left panel of Figure~\ref{fig:DMplot}, where we now vary $T_R$ for some fixed values of $m_a$. The right panel shows the various contributions to the relic abundance in dependence of $T_R$. }
    \label{fig:DMplot2}
\end{figure*}

Depending on the cutoff temperature $T_R$, the dominant dark matter production mechanism for the axion-Higgs portal is either $\gamma\gamma\to aa,\, b\bar b\to aa$ or $hh\to aa$, the transitions being at the scales $T_R\sim m_b$ and $T_R\sim m_h$, see Figure~\ref{fig:DMplot2}. In the large-$T_R$ limit, the process $hh\to aa$ dominates and an analytic calculation for the relic density is possible. For the relic density of axions which are heavy enough to be non-relativistic today, we find
\begin{align}
\begin{split}
    \frac{\Omega h^2}{0.12} &= \frac{2160}{\pi} \sqrt{\frac{10}{g_* g_{s*}^2}} \frac{sh^2}{0.12\,\rho_c } m_\text{Pl} m_a \Big(\frac{c_{ah}}{f^2}\Big)^2 T_R^3 \\
    &\approx 1.24\Big( \frac{m_a}{\SI{1}{eV}}\Big) \Big( \frac{c_{ah}/f^2}{10^{-14}\,\si{GeV^{-2}}}\Big)^2 \Big(\frac{T_R}{\SI{1}{TeV}}\Big)^3,
\end{split}
\end{align}
The dependence of the result on the effective number of degrees of freedom contributing to the energy density $g_*$ and the entropy density $g_{s*}$ is marginal, and therefore we evaluate them above the electroweak scale for all cases.
On the other hand, if the axion is sufficiently light i.e. $m_a\lesssim T_0$ with $T_0 = 2.3\cdot 10^{-4}\,\si{eV}$ the present temperature of the universe, we find instead
\begin{align}
\begin{split}
    \frac{\Omega h^2}{0.12} &= \frac{72\pi^3}{\zeta_3} \sqrt{\frac{10}{g_* g_{s*}^2}} \frac{sh^2}{0.12\rho_c} m_\text{Pl} T_0 \Big(\frac{c_{ah}}{f^2}\Big)^2 T_R^3 \\
    &\approx 1.97 \Big(\frac{c_{ah}/f^2}{5\cdot 10^{-13}\,\si{GeV^{-2}}}\Big)^2 \Big(\frac{T_R}{\SI{1}{TeV}}\Big)^3.
\end{split}
\end{align}

Searches for invisible Higgs decays therefore probe freeze-in dark matter through the axion-Higgs portal for low values of $T_R$. The window of cutoff temperatures for which the axion-Higgs portal can reproduce the observed relic density through freeze-in and that is probed by LHC searches is
\begin{align}
    \SI{300}{MeV} \lesssim T_R \lesssim \SI{2}{GeV}.
\end{align}
This window will shift to higher values of $T_R$ with improving precision on $c_{ah}/f^2$, as can be seen in Figure~\ref{fig:DMplot2}.

\subsection{Misalignment production}

For extremely small axion couplings, the observed relic abundance of DM cannot be obtained through either freeze-out or freeze-in production. If the axion is produced in a configuration away from its potential minimum in the early universe, it oscillates around the minimum at late times. The relic DM density is determined at the point $m_a = 3H$ where the oscillations of the axion field begin, and then diluted by cosmic expansion. We assume that the mechanism of spontaneous symmetry breaking generating the axion field happens after inflation, meaning that we have to average over the misalignment angle $\theta_0$ and find\footnote{We assume that the axion mass remains constant throughout the evolution of the universe. This is not valid in the case when the axion mass is generated by non-perturbative QCD effects, i.e. the QCD axion.} 
\begin{align}
\begin{split}
    \frac{\Omega h^2}{0.12} &= \frac{\pi^4}{30\sqrt{10}} \frac{h^2}{0.12\rho_c } \sqrt{g_{s*}} T_0^3 f^2 \sqrt{\frac{m_a}{m_\text{Pl}^3}}\\
    &\approx 0.37 \Big( \frac{m_a}{10^{-3}\si{eV}}\Big)^{1/2} \Big( \frac{f}{10^{13}\si{GeV}}\Big)^2.
\end{split}
\end{align}
Here $g_{s*}$ is the effective number of degrees of freedom contributing to the entropy density at the temperature when the axion field begins to oscillate.\footnote{Due to the mild dependence of $\Omega h^2$ on $g_{s*}$, we again evaluate $g_{s*}$ at the electroweak scale.}
Note that this result does not depend on the coupling $c_{ah}$, but only on the axion mass and the properties of the phase transition. 
There is no strict upper bound for the mass of axions which are produced through this mechanism. However, the coupling that is required to explain the observed DM relic density increases with the axion mass and at some point the working assumption of negligible thermal interactions will break down. We refrain from showing the parameter space for which the misalignment mechanism can reproduce the dark matter relic abundance in Figure~\ref{fig:DMplot} because it is independent of the coefficient $c_{ah}$. For values of $c_{ah}$ large enough to allow for freeze-in production the initial assumptions for misalignment are not fulfilled.

\subsection{Discussion}

The parameter space for which the observed relic DM abundance can be produced through the axion-Higgs portal is shown together with the strongest constraint from invisible Higgs decays in the left panel of  Figure~\ref{fig:DMplot}. For comparison the corresponding constraints on scalar Higgs portal dark matter is shown in the right panel of Figure~\ref{fig:DMplot}.

We find that the region where freeze-out production explains the observed amount of dark matter is always excluded for axions by the requirement $m_a\ll f$ and only allowed for $m_s > m_h/2$ for scalar dark matter.
The parameter space for which freeze-in production is allowed is shaded blue in Figure~\ref{fig:DMplot}. For larger axion masses, the temperature falls below the pair production threshold. For very low couplings, the effective theory becomes invalid for axions and the freeze-in production becomes independent of $T_R$ for scalars.
Freeze-in production is possible for axions or scalars with masses $m_a\gtrsim \SI{1}{eV}$ or $m_s\gtrsim \SI{1}{eV}$, below which dark matter is relativistic at the time of recombination independent of the temperature $T_R$. 

The sensitivity required for observing this type of dark matter increases with increasing axion mass and cutoff temperature. Currently, searches for invisible Higgs decays exclude values of $T_R\approx 0.3-2$ GeV. Any further improvement in invisible Higgs searches will probe dark matter production through the axion-Higgs portal for higher cutoff temperatures. 
In the left panel of Figure~\ref{fig:DMplot2} we show the contours of the correct relic abundance for different axion masses together with the constraint from invisible Higgs decays and the lower bound $T_R\gtrsim \SI{10}{MeV}$ set by BBN constraints. The blue shaded envelope indicates the range of cutoff temperatures $T_R$ that can be excluded for a given bound on invisible Higgs decays. The couplings necessary for DM production through the vacuum misalignment mechanism are so small that any confirmation of this scenario through dark matter production is not feasible. For the scalar Higgs portal the bound from supernova energy loss excludes values of the scalar Higgs coupling two orders of magnitude larger than the constraint from invisible Higgs decays. Light scalar dark matter is also constrained by bounds from Eot-Wash experiments sensitive to long-range forces induced by scalar exchange~\cite{Fichet:2017bng}.\footnote{We do not show the even stronger big bang nucleosynthesis bound from~\cite{Bauer:2020nld} in Figure~\ref{fig:DMplot}, because it only applies to wave-like dark matter which in turn could only be produced by vacuum misalignment.}

\section{Conclusions}
The axion-Higgs portal is the leading effective operator describing interactions between SM particles and axions or axion-like particles respecting the shift symmetry as well as a $Z_2$ symmetry. In the absence of any additional interaction the axion is stable and can only be produced in pairs. As a result, we find that very precise measurements such as searches for fifth forces do not result in relevant bounds, because the potential induced by the exchange of axion pairs scales as $V(r)\propto r^{-7}$ as a consequence of the derivative axion interaction in the axion-Higgs portal. Similarly, we show that the production of axions in the decays of pseudoscalar mesons $K\to \pi aa$, $B \to aa$ and vector mesons $V \to \gamma aa$ are suppressed by powers of the meson mass over the UV scale $f$. Bounds on the UV scale from atomic spectroscopy and meson decays are therefore substantially weaker compared to the bounds from searches for invisible Higgs decays. Invisible Higgs decays provide the strongest constraint on the axion-Higgs portal independent of the axion mass including astrophysical constraints from supernova cooling. An axion interacting through the axion-Higgs portal can provide an interesting dark matter candidate. While the production through freeze-out is excluded, searches for invisible Higgs decays can probe production through freeze-in for low reheating temperatures. Future $h \to $ invisible measurements can therefore discover freeze-in dark matter interacting through the axion-Higgs portal, the corresponding parameter space for the scalar Higgs portal is already excluded.  

\section{Acknowledgements}
We thank Joerg Jaeckel for pointing out to us the suppression of the axion-induced theta term in the case of the axion-Higgs portal. We further thank Robert Ziegler for his remarks about the contribution of $2\to 2$ processes for freeze-in production of dark matter. The work of Martin Bauer is supported by the Future Leader Fellowship DARKMAP.
\appendix

\section{Complex scalar UV completion\label{sec:UVcompletion}}

A very simple UV completion that induces am axion-Higgs portal is given by introducing a complex scalar singlet~\cite{Weinberg:2013kea}
\begin{align}\label{eq:ssbParametrization}
S=\frac{s+f}{\sqrt{2}}e^{ia/f}\,,    
\end{align}
with a vacuum expectation value $\langle S\rangle=f/\sqrt2$ and a Lagrangian
\begin{align}\label{eq:Slag}
\mathcal{L}_S=& \partial_\mu S \partial^\mu S^\dagger + \mu_s^2 S^\dagger S -\lambda_s (S^\dagger S)^2 -\lambda_{hs} S^\dagger S \phi^\dagger \phi\,\notag\\
&
+ \mathcal{L}_\text{SM}. 
\end{align}
This is equivalent to the assumption that no SM fields are charged under the global $U(1)$ associated with this complex scalar, such that linear interactions with $S$ are forbidden. Upon electroweak symmetry breaking, the radial mode $s$ mixes with the SM Higgs field. The dominant contribution to the mass of the radial mode is determined by the scale $f$, whereas the mass $m_a$ of the pseudo Nambu-Goldstone boson $a$ is generated by explicit symmetry breaking effects. In order for \eqref{eq:Slag} to match onto \eqref{eq:EFTLag} we assume that the scalar $s$ is sufficiently heavy and can be integrated out at the energy scales that we can access experimentally. From \eqref{eq:Slag} follows for the axion-Higgs portal
\begin{align}
c_{ah}= \frac{f}{v} \sin \alpha \qquad \text{with} \qquad \tan 2\alpha = \frac{2\lambda_{hs} vf}{m_s^2-m_h^2},
\end{align}
where $m_s$ and $m_h$ are the masses of the radial mode $s$ and the SM Higgs boson $h$, respectively. The coefficients in \eqref{eq:Slag} are constrained by measurements of the SM Higgs decays into SM particles. For example, a conservative estimate using bounds on the signal strength from LHC Higgs analyses leads to $|\sin\alpha| < 0.2$ for masses $m_s>m_h/2$~\cite{Dawson:2021jcl, ATLAS:2019nkf, CMS:2020gsy}. Further constraints on the parameters in \eqref{eq:Slag} arise from perturbativity, requiring a stable minimum of the potential and measurements of electroweak precision observables~\cite{Gonderinger:2012rd}.

\section{Relation to the strong CP problem\label{sec:strongCP}}

In addition to the well-established violation of the charge parity (CP) symmetry through weak interactions, the theory of quantum chromodynamics (QCD) allows for an additional operator violating CP symmetry 
\begin{align}\label{eq:thetaTerm}
    \mathcal{L}\ni \theta_c\frac{\alpha_s}{8\pi} G_{\mu\nu}^a \tilde G^{\mu\nu}_a,
\end{align}
where $\theta_c$ denotes the QCD theta angle. 
This parameter is experimentally found to be extremely small $\theta_c \lesssim 10^{-10}$~\cite{Baker:2006ts,Pendlebury:2015lrz,Graner:2016ses}. The question of why this coefficient is so small is called the strong CP problem.

The most famous solution to the strong CP problem is the QCD axion~\cite{Peccei:1977hh,Weinberg:1977ma,Wilczek:1977pj}. The QCD axion $a$ is a Goldstone boson of a spontaneously broken $U(1)$ symmetry which couples to the gluon field strength tensors
\begin{align}\label{eq:QCDaxion}
    \mathcal{L}\ni \frac{\alpha_s}{8\pi} \Big( \theta_c + \frac{a}{f} \Big) G^a_{\mu\nu} \tilde G^{\mu\nu}_a.
\end{align}
The coefficient of the $G\tilde G$ term is now a field which we call the effective QCD theta angle $\theta_\text{eff} = \theta_c + a/f$. At energies below the QCD scale, the coupling \eqref{eq:QCDaxion} generates an effective potential for the QCD axion which reads in the chiral limit $m_u=m_d$~\cite{GrillidiCortona:2015jxo}
\begin{align}\label{QCDaxionpotential}
    V_a = -\Lambda_\text{QCD}^4 \cos \theta_\text{eff}.
\end{align}
Assuming that other contributions to the QCD axion potential are negligible, one finds for the vacuum expectation value $\langle \theta_\text{eff} \rangle = 0$ or $\langle a\rangle = -f\theta_c$. 

The question whether the QCD contribution to the QCD axion potential dominates over other contributions is rather subtle and gives rise to another problem, the so-called QCD axion quality problem~\cite{Ghigna:1992iv,Kamionkowski:1992mf, Holman:1992us,Barr:1992qq}. It is argued that global symmetries like the $U(1)$ of the QCD axion should generically be explicitly broken at a high-energy scale $\Lambda_{U(1)}\lesssim m_\text{Pl}$, where $m_\text{Pl}$ is the Planck scale~\cite{Kallosh:1995hi, Witten:2017hdv}. In the effective theory, symmetry-breaking effective operators are then suppressed by powers of $\Lambda_{U(1)}$. 

For the case of the axion-Higgs portal, the coupling \eqref{eq:QCDaxion} is forbidden by the $Z_2$ symmetry prohibiting linear axion couplings. This $Z_2$ symmetry is a global symmetry and should therefore be explicitly broken at a scale $\Lambda_{Z_2}$ that need not be related to the scale of shift symmetry breaking $\Lambda_{U(1)}$ and should fulfill $\Lambda_{Z_2}\gg f$ in order to make the axion-Higgs portal a good effective theory, so that the equivalent to \eqref{eq:QCDaxion} reads
\begin{align}\label{eq:QCDaxion2}
    \mathcal{L}\ni \frac{\alpha_s}{8\pi} \Big( \theta_c + \frac{a}{\Lambda_{Z_2}} \Big) G^a_{\mu\nu} \tilde G^{\mu\nu}_a.
\end{align}
The axion couplings to Higgs bosons and gluons can therefore be controlled by different scales, but because the axion is periodic $a= a+2\pi f$, the allowed parameter space for the effective QCD theta angle becomes
\begin{align}
    \theta_\text{eff} = \theta_c + \frac{a}{\Lambda_{Z_2}} \in \Big[\theta_c -\pi \frac{f}{\Lambda_{Z_2}}, \theta_c + \pi \frac{f}{\Lambda_{Z_2}}\Big).
\end{align}
Given the hierarchy $f\ll \Lambda_{Z_2}$, the vacuum expectation value $\langle\theta_\text{eff}\rangle$ can only take values in the close vicinity of $\theta_c$. This is in contrast to the QCD axion, where the $Z_2$ symmetry is spontaneously broken as well, so that $\Lambda_{Z_2}=f$ and the axion field can balance an arbitrary value of $\theta_c$. We conclude that the mechanism that solves the strong CP problem for the QCD axion does not work if the axion-Higgs portal is the dominant interaction between SM fields and the axion. This discussion did not rely on the specific properties of the axion-Higgs portal, but holds for any axion model where the coupling \eqref{eq:QCDaxion} is forbidden by a global symmetry and only generated through explicit symmetry breaking. 

\section{Details about the freeze-in calculation\label{app:freezeIn}}

Detailed information about freeze-in production for the axion-Higgs portal is given in Figure~\ref{fig:DMplot2}. The relevant scales with decreasing cutoff temperature are $T_R\sim m_h$, where the dominant process changes from $hh\to aa$ to $b\bar b\to aa$, and $T_R\sim m_b$, after which the dominant process is $\gamma\gamma\to aa$. Due to the fact that freeze-in production in non-renormalizable $2\to 2$-processes is dominated at the highest temperatures and only possible for axions with masses below the cutoff temperature $m_a < T_R/2$, it is a good approximation to neglect the axion mass in the calculation of the yield $Y$.

\begin{figure*}
    \centering
    \includegraphics[scale=.58]{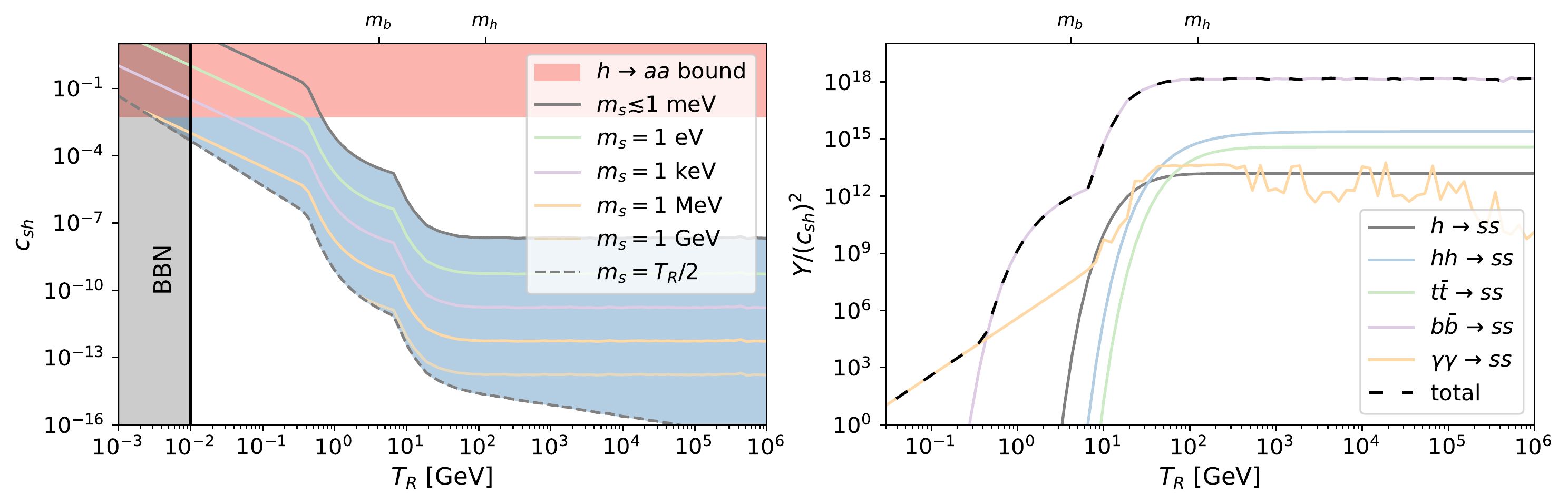}
    \caption{Like Figure \ref{fig:DMplot2}, but for the Higgs portal. The wiggles in the $\gamma\gamma\to ss$ line at high values of $T_R$ are due to numerical instabilities, however one can analytically show that this contribution is subleading in the limit of large cutoff temperatures. Depending on the value of the mass of the scalar $m_s$, stronger bounds on $c_{sh}$ might apply, this can be seen in figure~\ref{fig:DMplot}.}
    \label{fig:DMplot3}
\end{figure*}

The yield $Y$ has to be calculated numerically in the general case, but the result can be simplified in certain limits~\cite{Blennow:2013jba}. In the limit of large $T_R$, it is sufficient to neglect all particle masses and the result is just a power of the cutoff temperature. For the example of $hh\to aa$, one finds for the yield 
\begin{align}
    Y = \frac{2160}{\pi} \sqrt{\frac{10}{g_* g_{s*}}} \frac{c_{ah}^2 m_\text{Pl} T_R^3}{f^4}.
\end{align}
For processes with a Higgs mediator and cutoff temperatures around the Higgs mass, the dominant dark matter production comes from the pole of the Higgs boson propagator. For our example $b\bar b\to aa$, we can use $m_b\ll m_h$ and find 
\begin{align}
    Y = \frac{135}{4}\sqrt{\frac{10}{g_* g_{s*}^2}} \frac{c_{ah}^2 m_b^2 m_h^2 m_\text{Pl}}{f^4 \Gamma_h} \int_{m_h/T_R}^\infty dx x^3 K_1(x),
\end{align}
where $\Gamma_h$ is the decay width of the Higgs boson and $K_n(x)$ is the modified Bessel function of second kind. Finally, in the limit of small cutoff temperatures $T_R\ll m_h$, one can work with an effective theory where the Higgs boson is integrated out. For the process $\gamma\gamma\to aa$, we find 
\begin{align}
    Y = \frac{49766400}{7\pi} \sqrt{\frac{10}{g_* g_{s*}^2}} \frac{c_{ah}^2 c_\gamma^2 m_\text{Pl} T_R^7}{f^4 m_h^4}.
\end{align}
The relic density $\Omega h^2$ can be obtained from these results for the yield by using 
\begin{align}
    \Omega h^2 = \frac{sh^2}{\rho_c} \frac{\rho}{n} Y = \frac{sh^2}{\rho_c} Y \times \begin{cases}m_a\,, & m_a \gg T_0 \\ \frac{\pi^4}{30\zeta_3}T_0\,,& m_a \ll T_0\end{cases}.
\end{align}
Combining these analytical results, one can reconstruct the dominant contribution in the numeric solution shown on the right panel of Figure~\ref{fig:DMplot2} in most regions.

\section{Results for the Higgs portal}\label{app:Higgsportal}

Throughout the paper we compare the axion-Higgs portal with the scalar Higgs portal defined in \eqref{eq:higgsPortal} and we collect the relevant results for the scalar Higgs portal here.

The matching procedure outlined in Section~\ref{sec:aHportal} can be repeated in a straightforward way with the replacements $\partial_\mu \to 1$ and $c_{ah}/f^2\to c_{sh}$, leaving the Wilson coefficients unchanged. For the Higgs decay width into scalars we find 
\begin{align}
    \Gamma ( h\to ss ) = \frac{v^2}{8\pi m_h} c_{sh}^2 \sqrt{1-4\frac{m_s^2}{m_h^2}}.
\end{align}
For the flavor-changing meson decays, we find
\begin{align}
    \Gamma (B^0\to ss ) &= \frac{m_{B^0}^3}{32\pi} \frac{c_{sh}^2 c_{bd}^2 f_{B^0}^2}{m_h^4} \sqrt{1-4\frac{m_s^2}{m_{B^0}^2}}, \\
    \Gamma (K^+\to \pi^+ ss) &=\frac{m_{K^+}^5}{2^{10}\pi^3} \frac{c_{sh}^2 c_{ds}^2}{m_h^4} K  \Big( \frac{m_s^2}{m_{K^+}^2}, \frac{m_{\pi^+}^2}{m_{K^+}^2}\Big),
\end{align}
where the function $K (a,b)$ is defined as
\begin{align}
    K (a,b) = 4(1-b)^2 &\int_{4a}^{(1-\sqrt{b})^2} dx \sqrt{x-4a}\notag\\&\times \sqrt{\Big( \frac{1-b-x}{2\sqrt{x}}\Big)^2-b}.
\end{align}
Finally, for radiative quarkonia decays, we find
\begin{align}
\begin{split}
    &\mathcal{R} ( V\to \gamma ss) \equiv \frac{\Gamma ( V\to \gamma ss)}{\Gamma (V\to e^+e^-)} \\
    &= \frac{1}{2^7 \pi^3\alpha} \frac{c_{sh}^2 
    m_V^4}{m_h^4} K\Big(\frac{m_s^2}{m_V^2}, 0\Big) \bigg[1-\frac{4\alpha_s}{3\pi}a_H(z)\bigg],
\end{split}
\end{align}
with the function
\begin{align}
\begin{split}
    K(x, 0) =& \sqrt{1-4x} (1+2x) \\ &-4(1-x)x\log\frac{1+\sqrt{1-4x}}{1-\sqrt{1-4x}}
\end{split}
\end{align}

For scalar Higgs portal dark matter as discussed in Section~\ref{sec:darkmatter} we obtain for the yield from freeze-out production after the QCD phase transition
\begin{align}
\begin{split}
    \frac{\Omega h^2}{0.12} &= 120 \frac{sh^2}{ 0.12\rho_c} \frac{m_h^4}{c_{sh}^2 c_\gamma^2 m_a^2 m_\text{Pl}} \\
    &\approx 0.50 \Big( \frac{c_{sh}}{5\cdot 10^4}\Big)^{-2} \Big( \frac{m_a}{\SI{100}{MeV}}\Big)^{-2}\,,
\end{split}
\end{align}
and for freeze-out before the QCD phase transition
\begin{align}
\begin{split}
    \frac{\Omega h^2}{0.12} &= 24\sqrt{10} \frac{sh^2}{0.12\rho_c} \frac{m_h^4}{c_{sh}^2 m_b^2 m_\text{Pl}}\\
    &\approx 1.13 \Big( \frac{c_{sh}}{10^0}\Big)^{-2}\,.
\end{split}
\end{align}
 In contrast to the case of the axion-Higgs portal, there is still a small window for freeze-out production at large values of the axion mass. The corresponding limits for the dark matter mass and the relation for the freeze-out temperature $x_\text{FO} = m_a/T_\text{FO}$ are both independent of the specific process, and therefore one finds the same results as for the axion-Higgs portal. 
 
Freeze-in production through the scalar Higgs portal is very different from the axion-Higgs portal, because the former is a renormalizable operator. This means that dark matter production won't be most effective at the highest temperatures, but at poles or thresholds. Therefore, above a certain value of $T_R$ the parameter space for which freeze-in production results in the correct relic abundance is independent of $T_R$. This results in a lower bound on the coupling $c_{sh}$ indicated by the lower limit of the blue region on the right panel of in Figure~\ref{fig:DMplot}.

It turns out that the naive production processes $h\to ss$ and $hh\to ss$ are always subleading, whereas the dominant contribution comes from $\gamma\gamma\to ss$ or $b\bar b\to ss$, depending on the cutoff temperature $T_R$. This can be seen in the right panel of Figure~\ref{fig:DMplot3} which contains detailed information about freeze-in production through the Higgs portal in analogy to Figure~\ref{fig:DMplot2} for the axion-Higgs portal. 

Similar to Appendix~\ref{app:freezeIn}, we give expressions for the yield from the leading processes in special limits. For cutoff temperatures above the electroweak scale, the process $b\bar b\to ss$ dominates. The corresponding yield reads 
\begin{align}
\begin{split}
    Y &= \frac{270}{\pi} \sqrt{\frac{10}{g_* g_{s*}^2}} c_{sh}^2 \frac{m_b^2 m_\text{Pl}}{m_h^2 \Gamma_h} \int_{m_h/T_R}^\infty dx x^3 K_1(x) \\
&= 405 \sqrt{\frac{10}{g_* g_{s*}^2}} c_{sh}^2 \frac{m_b^2 m_\text{Pl}}{m_h^2 \Gamma_h}\qquad \text{for}\qquad T_R\gg m_h.
\end{split}
\end{align}
For cutoff temperatures sufficiently far below the Higgs mass, the process $\gamma\gamma\to ss$ can be described by an effective field theory where the Higgs boson is integrated out. This yields 
\begin{align}
    Y = \frac{17280}{\pi} \sqrt{\frac{10}{g_* g_{s*}^2}} c_{sh}^2 c_\gamma^2 \frac{m_\text{Pl} T_R^3}{m_h^4}.
\end{align}
Further, for the scalar Higgs portal bounds from supernova cooling and Eot-Wash experiments become dominant for lower axion masses, as shown in Figure~\ref{fig:DMplot}.


\end{document}